\documentclass[twocolumn,showpacs,prb]{revtex4}

\usepackage{amssymb}
\usepackage{graphicx}
\usepackage{graphics}
\usepackage{amsmath}

\begin{document}

\title{Dielectric response due to stochastic motion of pinned domain walls}
\author{A. A. Fedorenko, V. Mueller, and S. Stepanow}
\affiliation{Martin-Luther-Universit\"{a}t Halle, Fachbereich Physik, D-06099\\
Halle, Germany}
\date{\today }
\pacs{05.20.-y, 74.25.Qt, 64.60.Ak, 75.60.Ch}

\begin{abstract}
We study the contribution of stochastic motion of a domain wall (DW) to the
dielectric AC susceptibility for low frequencies. Using the concept of
waiting time distributions, which is related to the energy landscape of the
DW in a disordered medium, we derive the power-law behavior of the complex
susceptibility observed recently in some ferroelectrics below Curie
temperature.
\end{abstract}

\maketitle

During the last decade considerable interest has been attracted to the
anomalously high dielectric susceptibility observed in several
ferroelectrics below the paraelectric-ferroelectric phase transition
temperature $T_{c}$. The theory of critical phenomena predicts in the
vicinity of $T_{c}$ the Curie-Weiss law for the temperature dependent static
susceptibility. However, for some ferroelectrics, such as potassium
dihydrogen phosphate\cite{mueller1} ($\mathrm{KH_{2}PO_{4}}$) or rubidium
dihydrogen phosphate\cite{mueller2} ($\mathrm{RbH_{2}PO_{4}}$), the complex
susceptibility $\chi (\omega )=\chi ^{\prime }(\omega )-i\chi ^{\prime
\prime }(\omega )$ within the so-called plateau range $T_{f}<T<T_{c}$
remains unusually high. This anomaly is usually ascribed to DW motion, an
activity which freezes out below $T_{f}$. According to Ref.~%
\onlinecite{mueller1}, the low-frequency dielectric spectrum in the plateau
range of $\mathrm{KH_{2}PO_{4}}$ consists of non-Debye contribution, which
has a power-law behavior with a small exponent, and several Debye-like
constituents. Recent experiments performed on $\mathrm{%
RbFe_{1/2}Nb_{1/2}O_{3}}$ \cite{park} and $\mathrm{%
Sr_{0.61-x}Ce_{x}Ba_{0.39}Nb_{2}O_{6}}$ (\textit{SBN}:Ce) \cite{kleemann02}
revealed in the range of low frequency $\omega <\omega _{c}$ a power-law
behavior of the DW response
\begin{equation}
\chi (\omega )=\chi _{\infty }(1+1/{(i\omega \tau _{0})^{n}})  \label{eq2}
\end{equation}%
with $0<n<1$. Here $\omega _{c}$ is the threshold frequency separating the
low- and high- frequency regimes, $\tau _{0}$ the characteristic relaxation
time and $\chi _{\infty }$ the high-frequency limit of the complex
susceptibility. The real and imaginary parts of the complex susceptibility (%
\ref{eq2}) are related by equation
\begin{equation}
\chi ^{\prime \prime }(\omega )=(\chi ^{\prime }(\omega )-\chi _{\infty
})\tan (n\pi /2),  \label{susc-real}
\end{equation}%
which follows from the Kramers-Kronig relation. In a range of frequencies
above $\omega _{c}$, the imaginary part of susceptibility $\chi ^{\prime
\prime }(\omega )$ was found to increase linearly on a log-log scale, while
the corresponding real part $\chi ^{\prime }(\omega )$ decreases linearly on
a linear-log scale. \cite{kleemann02}

An attempt to explain phenomenologically the existence of two regimes of the
dynamic response was made in Ref.~\onlinecite{chen02}, where the
low-frequency behavior (\ref{eq2}) was attributed to the irreversible
viscous motion of DWs in a disordered medium. However, this interpretation,
based on the picture of the dynamic hysteresis developed in Ref.~%
\onlinecite{nattermann01}, is restricted to the adiabatic regime only.

Eqs.~(\ref{eq2}) and (\ref{susc-real}) are in accordance with Jonscher's
universal dielectric response law, \cite{jonscher-nature,jonscher} which
describes the response of a wide class of dielectrics in a broad range of
frequencies. Generally, this behavior is related to hopping of charge
carriers, such as electrons, polarons or various ions, between different
localized states. The exponent $n$ is determined by the spatial and
energetic distribution of the corresponding states. The ferroelectric
non-Debye dielectric response (\ref{eq2}) is expected to be due to jumps of
DWs (or DW segments) between different metastable states.

In the present paper, we study the contribution of stochastic motion of a DW
to the dielectric susceptibility at low frequencies, using the concept of
waiting time distribution, which is related to the energy landscape of the
DW in a disordered medium. We restrict our consideration to the regime in
which the field driven motion of adjacent DWs is independent of each other.
According to Ref.~\onlinecite{kleemann-ferro} the exponent $n$ depends on
the poling state of dielectric samples. Poling is expected to decrease the
density of DWs, which would improve the single DW approach. However, among
others it may also change the distribution of impurities.

The DW, which separates domains of different polarization, can be viewed as
a two-dimensional ($d=2$) elastic interface moving in a disordered
environment. The configuration of a $d$-dimensional elastic interface is
described by the profile function $z(x,t)$, which obeys the Langevin-type
equation
\begin{equation}
\mu ^{-1}\frac{\partial z(x,t)}{\partial t}=\gamma \nabla ^{2}z+h_{0}\cos
\omega t+g(x,z)+\eta (x,t),  \label{motion}
\end{equation}%
where $\mu $ is the mobility, $\gamma $ the stiffness constant and $%
h(t)=h_{0}\cos \omega t$ the AC driving force density. Note that $x$ is the $%
d$-dimensional vector. We assume Gaussian distribution of thermal noise,
with zero mean and the correlator
\begin{figure}[tbph]
\includegraphics[clip,width=2.5in]{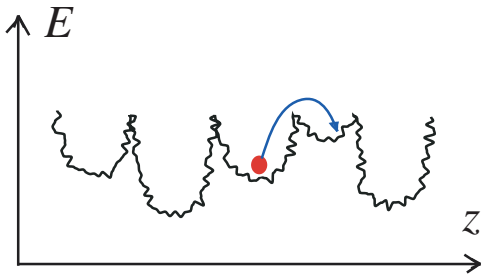}
\caption{The energy landscape of the center of mass of DW }
\label{fig1}
\end{figure}
\begin{equation}
\overline{\eta (x,t)\eta (x^{\prime },t^{\prime })}=2\mu ^{-1}T\delta
^{(d)}(x-x^{\prime })\delta (t-t^{\prime }).  \label{thermal}
\end{equation}%
Note that in real ferroelectrics the dependence of $\mu $ and $\gamma $ on
temperature $T$ may complicate the comparison of the computed temperature
dependence of the susceptibility with experiment. Pinning to impurities,
which are described by the quenched random force $g(x,z)$, strongly
suppresses the DW mobility. The quenched forces are assumed to be Gaussian
distributed with zero mean and the variance
\begin{equation}
\left\langle g(x,z)g(x^{\prime },z^{\prime })\right\rangle =\delta
^{(d)}(x-x^{\prime })\Delta (z-z^{\prime }).  \label{disorder}
\end{equation}%
To make this model well-defined, we introduce the cutoff $\Lambda _{0}^{-1}$
in the Dirac's delta function $\delta ^{d}(x)$ at scales in the order of the
impurity separation. In what follows we consider random field (RF) disorder,
where for $z>0$ the correlator $\Delta (z)=\Delta (-z)$ is a monotonically
decreasing function of $z$, and decays rapidly to zero over the distance $a$.

To describe different regimes of the DW motion, we recall the fundamental
scales of the problem. On scales smaller than the Larkin scale $L_{c}\simeq
(a^{2}\gamma ^{2}/\Delta (0))^{1/\varepsilon }$, the elasticity wins over
the disorder for $d<4$ and the DW remains flat, while on larger scales the
DW becomes rough and self-affine with a nontrivial roughness exponent $\zeta
$. \cite{fisher86} At $T=0$ and $\omega \rightarrow 0$, the system (\ref%
{motion})-(\ref{disorder}) undergoes a second order depinning transition at $%
h_{0}=h_{c}$. \cite{nstl92,narayan93,chauve01} The depinning threshold $%
h_{c}\simeq a\gamma L_{c}^{-2}$ can be estimated from balancing the pinning
forces and the driving forces on the scale $L_{c}$. At finite temperature,
the depinning transition smears out, and the DW moves at any field strength.
In the creep regime, which sets in for low fields, the motion is very slow
and controlled by thermal activation. The energy barrier $E_{B}(L)\simeq
U_{c}(L/L_{c})^{\theta }$ has to be overcome in order to move a DW segment
of linear size $L$, where $\theta =2\zeta +d-2$, and $U_{c}\simeq
a^{2}\gamma L_{c}^{d-2}$ is the typical barrier on the Larkin scale $L_{c}$.
The DW segment of size $L$ spends time $\tau (L)=\tau _{0}\exp (E_{B}(L)/T)$
in a valley before it jumps to the next valley. Here $\tau
_{0}=L_{c}^{2}/(\mu \gamma )$ is the microscopic time. For low fields $h\ll
h_{c}$, the average DW velocity is determined by the thermal activation of
segments with optimal size $L_{\mathrm{opt}}=L_{c}(h/h_{c})^{-1/(2-\zeta )}$,
and obeys the creep law
\begin{equation}
v\propto \exp [-U_{c}/T(h/h_{c})^{-\mu }],  \label{creep}
\end{equation}%
with $\mu =(d-2+2\zeta )/(2-\zeta )$.

If the DW is driven by AC field with the frequency $\omega $, the barriers
with waiting times $\omega \tau (L)>1$ cannot be overcome during one period
of the field. Thus the threshold frequency
\begin{equation}
\omega ^{\ast }=1/\tau (L_{\mathrm{opt}})=\tau _{0}^{-1}\exp
(-U_{c}/T(h_{0}/h_{c})^{-\mu })  \label{h-T-threshold}
\end{equation}%
separates the sliding regime from the pinned regime.\cite{nattermann01} The
critical value $\omega ^{\ast }$, which corresponds to the experimental
setup used in Refs.~\onlinecite{kleemann02} and \onlinecite{kleemann-ferro},
can be estimated as follows. Since the activity of DWs for $T<T_{f}$ is
almost frozen, \cite{huang97} we conclude that temperatures in the range $%
T_{f}<T<T_{c}$ are of order $U_{c}$. The creep exponent is $\mu (d=2)=1$ for
RF disorder. The amplitude of the electric field in the dielectric
experiments of Ref.~\onlinecite{kleemann-ferro} $h_{0}\approx 300V/m$ is far
below the coercive field $150kV/m$. Assuming that the threshold $h_{c}$ is
of order of the coercive field we find that the experimental frequencies ($%
\omega >0.01$Hz) are likely to be much higher than the threshold frequency
estimated from Eq.(\ref{h-T-threshold}). To explain the observed
low-frequency response we suppose that DWs can also move for $\omega ^{\ast
}<\omega <\omega _{c}$ ($\omega _{c}$ is defined below) via hopping between
different minima of the energy landscape. However, this motion is extremely
slow and can be characterized by the displacement $z\propto t^{n}$ with $n<1$%
, \cite{feigelman88} so that in the adiabatic limit $t\rightarrow \infty $
the average velocity will vanish. In analogy with the stochastic transport
phenomena in disordered solids, \cite{scher73} we refer to the DW-hopping in
the frequency range $\omega ^{\ast }<\omega <\omega _{c}$ as the stochastic
regime.

To calculate the dielectric response in the stochastic regime, we consider a
$180^{\circ }$ domain structure of quasiperiodicity $2l$, where the DW
separates homogeneously polarized regions with spontaneous polarization $%
P_{0}$ and $-P_{0}$, respectively. The time dependence of the macroscopic
polarization reads
\begin{equation}
dP(t)/dt=(2P_{0}/l)\langle v(t)\rangle .  \label{P1}
\end{equation}%
The linear susceptibility $\chi (\omega )$ defined through the Fourier
transform $P(\omega )=\chi (\omega )h(\omega )$ of $P(t)$ can be expressed
as
\begin{equation}
\chi (\omega )=\frac{2P_{0}}{l}\frac{\mu (\omega )}{i\omega },  \label{chi11}
\end{equation}%
where $\mu (\omega )$ is the renormalized DW mobility given by $\langle
v(\omega )\rangle =\mu (\omega )h(\omega )$. The center of mass of the field
driven DW probes different local minima of the rugged energy landscape,
corresponding to different metastable DW-configurations (see
Fig.~\ref{fig1}). This motion is similar to the stochastic motion of a
particle in the RF environment.\cite{feigelman88} The Langevin equation for
the particle is given by
\begin{equation}
\frac{dz}{dt}=f(z)+h+\eta (t),
\end{equation}%
where $f(z)$ is the Gaussian distributed RF with $\langle f(z)\rangle =0$, $%
\langle f(z)f(z^{\prime })\rangle =\Delta _{0}\delta (z-z^{\prime })$, and $%
\eta (t)$ the thermal noise. At low temperatures $T$ and small driving
forces $h$, the motion of the particle can be described by the thermally
activated dynamics, which is controlled by the probability distribution
density $W(E)$ of energy barriers $E$ separating different metastable
states. Using the standard methods of the field theory, $W(E)$ can be
written as a path integral, the estimate of which by the steepest descent
method gives \cite{feigelman88}
\begin{equation}
W(E)=E_{0}^{-1}\exp (-E/E_{0})  \label{E0}
\end{equation}%
with $E_{0}=\Delta _{0}/(2h)$. Due to the thermally activated dynamics, the
waiting time $\tau (E)=\tau _{0}\exp (-E/T)$ needed to overcome the barrier $%
E$ is characterized by the distribution $\Psi (\tau )d\tau =W(E)dE$, which
decays as a power law for large $\tau $
\begin{equation}
\Psi (\tau )=\frac{n\tau _{0}^{n}}{\tau ^{1+n}},\ \ \ n=T/E_{0}.
\label{psi0}
\end{equation}%
As follows from Eq.~(\ref{psi0}) the average waiting time $\int d\tau \tau
\Psi (\tau )$ diverges for low temperatures $T<E_{0}$. This indicates that
the particle motion is dominated by very rare, but extremely deep potential
wells, which leads to sublinear drift $z\propto t^{n}$, $n<1$.

The distribution of waiting times which controls the motion of the center of
mass of DW can also be related to the distribution of the energy barriers $%
W(E)$. \cite{vinokur96} As shown in Ref.~\onlinecite{drossel-kardar95}, the
energy barriers scale as the energy minima, and therefore, have identical
distributions. Generalizing the loop expansion of the energy distribution
developed in Ref.~\onlinecite{our} for driven elastic interface we have
obtained the energy distribution of the DW in the presence of nonzero
external force $h$ as
\begin{eqnarray}
&&W(E)=\int_{-\infty }^{\infty }\frac{ds}{2\pi }\exp (isE)\exp \left[ -\frac{%
1}{2}L^{d}\right.  \notag \\
&&\mbox{}\hspace{4mm}\times \int_{k}\ln \left( 1+\frac{is}{\gamma k^{2}}%
\left. \int_{q}\frac{\Delta _{q}}{1+q^{2}h^{2}/(\gamma ^{2}k^{4})}\right) %
\right] ,  \label{e14}
\end{eqnarray}%
where $\Delta _{q}$ is the Fourier transform of $\Delta (z)$. Assuming that
the width $a$ of the disorder correlator $\Delta _{q}$ is small, we
approximate the integral over $q$ in Eq.~(\ref{e14}) by $\Delta
(0)/(1+h^{2}a^{2}/\gamma ^{2}k^{4})$. The inverse Fourier transform of the
discrete version of Eq.~(\ref{e14}) for $d=1$ gives $f(x;\kappa
)=\sum\nolimits_{j=1}^{\infty }C_{j}(\kappa )\exp (-(j^{2}+\kappa
^{2}/j^{2})x)$, where $x=4\pi ^{2}\gamma E/(\Delta (0)L^{2})$, $\kappa
=(haL^{2})/(4\pi ^{2}\gamma )$, and the coefficients $C_{j}(\kappa )$ are
given by
\begin{eqnarray}
C_{j}(\kappa ) &=&\frac{2j^{2j-5}}{\pi ^{2}{\kappa }}\Gamma (2j)\Gamma
\left( j-\frac{{\kappa }}{j}\right) \Gamma \left( j+\frac{{\kappa }}{j}%
\right)  \notag \\
&\times &(j^{4}-\kappa ^{2})\frac{\cosh ^{2}(\pi \sqrt{\frac{\kappa }{2}}%
)-\cos ^{2}(\pi \sqrt{\frac{\kappa }{2}})}{\prod\limits_{l=1}^{j-1}[(l^{4}+%
\kappa ^{2})j^{2}-l^{2}(j^{4}+\kappa ^{2})]}.  \label{e26}
\end{eqnarray}%
Using the steepest descent method to estimate the sum in $f(x;\kappa )$, we
find that for large energy $E$ the distribution is exponential $W(E)\propto
\exp (-2ahE/\Delta (0))$. However, this expression of the energy
distribution is incorrect for the same reason as the perturbation result for
the roughness. \cite{our} The appropriate method to extract the correct
behavior of elastic interfaces in random environment is the functional
renormalization group. \cite{fisher86,nstl92,narayan93,chauve01,our} To find
the effect of renormalization on the energy distribution for driven
interface, we follow Ref.~\onlinecite{our} and replace all bare quantities
in Eq.~(\ref{e14}) by the renormalized ones. We obtain the renormalized
energy distribution for $d=1$ as $W(E)\propto \exp (-2(\gamma
ah)^{1/2}E/\Delta (0)L_{c})$. The evaluation of Eq.~(\ref{e14}) for $d>1$ is
more complicated. However, for $d=2$ the tail of the energy distribution for
large energies is exponential, and is given by Eq.~(\ref{E0}) with $E_{0}$
of order $U_{c}$. Consequently, the distribution of waiting times has the
same form (\ref{psi0}).

To calculate the renormalized mobility $\mu (\omega )$, we now employ the
Montroll-Weiss formalism developed in Refs.~\onlinecite{scher73} and %
\onlinecite{montroll65} for continuous time random walks. According to Ref.~%
\onlinecite{scher73}, the mobility $\mu (\omega )$ can be written as
\begin{eqnarray}
\mu (\omega ) &=&-\frac{\omega ^{2}}{2T}\int\limits_{0}^{\infty
}dte^{-i\omega t}\langle (z(t)-z(0))^{2}\rangle  \notag \\
&=&\frac{\sigma _{0}^{2}}{2T}\frac{i\omega \tilde{\Psi}(i\omega )}{1-\tilde{%
\Psi}(i\omega )},  \label{eq18}
\end{eqnarray}%
where $\sigma _{0}$ is the average distance between adjacent potential wells
and $\tilde{\Psi}(u)$ is the Laplace transform of $\Psi (\tau )$. For $\Psi
(\tau )$ given by Eq.~(\ref{psi0}) we obtain
\begin{equation}
\tilde{\Psi}(u)=\int\limits_{0}^{\infty }dte^{-u\tau }\Psi (\tau )\simeq
1-A(u\tau _{0})^{n},  \label{eq19}
\end{equation}%
where $A=\pi /(\Gamma (n)\sin (\pi n))$. Combining Eqs.~(\ref{chi11}), (\ref%
{eq18}), and (\ref{eq19}), we obtain the complex susceptibility as
\begin{equation}
\chi (\omega )=\frac{2P_{0}}{Al}\frac{\sigma _{0}^{2}}{2T}\frac{1}{(i\omega
\tau _{0})^{n}}.  \label{res1}
\end{equation}

The above consideration applies for low frequencies, where the center of
mass of the DW explores many valleys during one period $2\pi /\omega $ . For
$\omega $ larger than some frequency $\omega _{c}$, the center of mass of
the DW stays in one valley. The characteristic frequency $\omega _{c}$ can
be estimated from the condition that the center of mass of the DW remains
within one well during period $2\pi /\omega $ with the probability $1$. This
yields $\omega _{c}\approx \tau _{0}^{-1}$.

For frequencies $\omega >\omega _{c}$, the DW is captured in one valley, the
potential of which we approximate by
\begin{equation}
U(\bar{z})=\frac{1}{2\tau _{1}}\bar{z}^{2}+\int\limits_{0}^{\bar{z}%
}dz^{\prime }\bar{g}(z^{\prime }),  \label{eqUz}
\end{equation}%
where the 1st term describes the average shape of the well, and the 2nd term
describes the fluctuations due to the residual part of the random force $g$.
Neglecting the second term in Eq.(\ref{eqUz}) yields Debye-response
\begin{equation}
\chi ^{\prime \prime }(\omega )\propto \omega \tau _{1}/(1+(\omega \tau
_{1})^{2}).  \label{deb-large}
\end{equation}%
\begin{figure}[tbph]
\includegraphics[clip,width=3.5in]{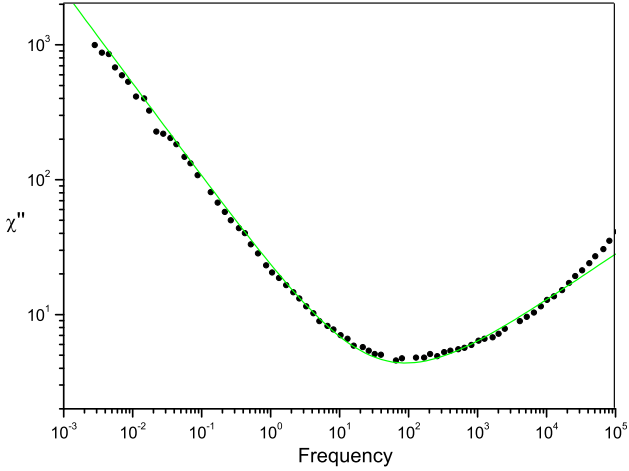}
\caption{The imaginary part of the susceptibility of poled \textit{SBN}:Ce
measured at $T=294$ K. \protect\cite{kleemann-ferro} The best fit by Eq.~(%
\protect\ref{ansatz}) corresponds to $a=21.56$, $n=0.69$, $\protect\kappa %
=(n+n_{1})/2n=0.75$ ($n_{1}=0.35$), and $\protect\omega _{c}=91$.}
\label{fig2}
\end{figure}
The 2nd term in Eq.~(\ref{eqUz}) results in distribution of relaxation times
$\Psi _{1}(\tau )$ instead of the single relaxation time $\tau _{1}$.
Expecting the exponentially distributed barriers on all scales, the
distribution of relaxation times is likely to obey a power-law $\Psi
_{1}(\tau )=n_{1}\tau _{1}^{n_{1}}/\tau ^{1+n_{1}}$ with some exponent $%
n_{1} $ which is not necessarily equal to $n$. As a result the Debye
response (\ref{deb-large}) changes to
\begin{equation}
\chi ^{\prime \prime }(\omega )\propto (\omega \tau _{1})^{n_{1}}.
\label{res2}
\end{equation}%
To describe $\chi ^{\prime \prime }(\omega )$ for $\omega $ $\lessgtr \omega
_{c}$ we propose the following scaling ansatz
\begin{equation}
\chi ^{\prime \prime }(\omega )=(\omega \tau _{1})^{n_{1}}F(L_{\omega
}/\sigma _{0}).  \label{eq17}
\end{equation}%
$F(x)$ is a scaling function and $L_{\omega }$ is the distance in $z$
direction explored by the DW during one period of the AC field. The distance
$L(\tau _{\mathrm{m}})$ explored by the DW during the time $\tau _{\mathrm{m}%
}$ can be estimated from the condition
\begin{equation}
\frac{L(\tau _{\mathrm{m}})}{\sigma _{0}}\int\limits_{\tau _{\mathrm{m}%
}}^{\infty }\Psi (\tau )d\tau =1,  \label{L-o}
\end{equation}%
which means that the DW at most once is captured by the deepest well with
the waiting time larger than $\tau _{\mathrm{m}}$. For $\tau _{\mathrm{m}}$
equal to the period of AC field, Eq.(\ref{L-o}) gives $L_{\omega }\equiv
L(\tau _{\mathrm{m}}\simeq 1/\omega )=\sigma _{0}/(\omega \tau _{0})^{n}$.
Substituting the latter into Eq.~(\ref{eq17}) we obtain
\begin{equation}
\chi ^{\prime \prime }(\omega )=(\omega \tau _{1})^{n_{1}}F(1/(\omega \tau
_{0})^{n}).
\end{equation}%
According to Eqs.(\ref{res1}) and (\ref{res2}), the scaling function $F(x)$
must behave for small and large $x$ as%
\begin{equation*}
F(x)\simeq \left\{
\begin{array}{cc}
x^{n_{1}}, & x\ll 1, \\
x^{1+n_{1}/n}, & x\gg 1.%
\end{array}%
\right.
\end{equation*}%
The fit of the experimental data of Kleemann \textit{et al.,}~\cite%
{kleemann-ferro} using the scaling function%
\begin{equation}
\chi ^{\prime \prime }(\omega )=a\omega ^{-n}\left[ 1+\frac{1}{2\kappa -1}%
\left( \frac{\omega }{\omega _{c}}\right) ^{\kappa }\right] ^{2n}.
\label{ansatz}
\end{equation}%
is shown in Fig.~\ref{fig2}.

In contrast to the imaginary part, the real part of the susceptibility
behaves as a power of $\ln \omega $. \cite{kleemann02,kleemann-ferro} To
explain this behavior, we suggest that the intrawell dynamics of DW is a
superposition of center of mass motion and relaxational motion of internal
modes. Indeed, in the pinned phase $\omega >\omega _{c}$, the DW segments
can also jump between metastable states with close energies, which gives
rise to additional dissipation. The relaxational contribution of internal
modes to the susceptibility reads \cite{ioffe,nattermann90}
\begin{eqnarray}
&&\chi ^{\prime }\propto \left( \ln (1/{\omega \tau _{0}})\right) ^{2/\theta
},  \label{nt1} \\
&&\chi ^{\prime \prime }\propto \left( \ln (1/{\omega \tau _{0}})\right)
^{2/\theta -1}.  \label{nt2}
\end{eqnarray}%
We briefly remind the derivation of Eqs.~(\ref{nt1}) and (\ref{nt2}). As
suggested in Refs.~\onlinecite{ioffe} and \onlinecite{nattermann90}, one can
treat the pinned DW as an hierarchical ensemble of noninteracting two-level
systems (TLS), each is a DW segment of linear size $L$. The separation
between the two configurations of the given TLS is $w(L)\propto L^{\zeta }$,
and the energy difference is $\Delta E$. Transitions caused by thermal
activation may occur only in TLSs with the energy difference $\Delta E\leq T$%
. The corresponding rate of transitions is given by $\tau ^{-1}(L)=\tau
_{0}^{-1}\exp (-E_{B}(L)/T)$, where $E_{B}(L)$ is the energy barrier. In the
presence of AC field, these transitions lead to power dissipation \cite%
{ioffe}
\begin{equation}
Q(L)\simeq \frac{(\delta E(L))^{2}}{4T\cosh ^{2}({\Delta }/{2T})}\frac{%
\omega ^{2}\tau (L)}{1+\omega ^{2}\tau ^{2}(L)},  \label{Q1}
\end{equation}%
where $\delta E=P_{0}L^{d}w(L)h_{0}$ is the shift of $\Delta E$ due to the
external field $h$. To obtain the power dissipation density, we have to
average Eq.~(\ref{Q1}) over the distribution $P_{L}(\Delta E)$, and then sum
over all TSLs in the unit volume. Since the distribution function $%
P_{L}(\Delta E)$ is expected to be smooth with a width of the order $\Delta
E(L)\propto L^{\theta }\gg T$, the mean value of the factor $\cosh ^{-2}({%
\Delta E}/{2T})$ gives the fraction of the thermally active TLSs $T/\Delta
E(L)$. The hierarchical structure of TLSs implies that the density of TLSs
of linear size $L$ is $1/L^{d+1}$. Therefore the power dissipation density $%
\overline{Q}$ can be written as
\begin{equation}
\overline{Q}\propto P_{0}h_{0}^{2}\int\limits_{L_{c}}^{\infty }dLL\frac{%
\omega ^{2}\tau (L)}{1+\omega ^{2}\tau ^{2}(L)}.  \label{Q2}
\end{equation}%
Due to the exponential increase of $\tau $ with $L$, the main contribution
to the integral (\ref{Q2}) comes from the length scale $\overline{L}_{\omega
}$ given by the condition $\omega \tau (\overline{L}_{\omega })\sim 1$. On
the other hand, the power dissipation is related to the imaginary part of
the complex susceptibility \cite{landau}
\begin{equation}
\overline{Q}=\frac{1}{2}\omega \chi ^{\prime \prime }h_{0}^{2}.  \label{P0}
\end{equation}%
Equating (\ref{Q2}) and (\ref{P0}) we arrive at Eq.~(\ref{nt2}) for the
imaginary part of the susceptibility. The corresponding real part (\ref{nt1}%
) can be restored from the imaginary part (\ref{nt2}) with the help of the
Kramers-Kronig relation in the form of "$\pi /2$ rule" \cite{pytte87}
\begin{equation}
\chi ^{\prime \prime }(\omega )=-\frac{\pi }{2}\frac{d\chi ^{\prime }(\omega
)}{d\ln \omega }.
\end{equation}

The real part of the susceptibility, which corresponds to Eq.~(\ref{res2}),
decreases as a power-law and can be hidden by the slower decreasing
logarithmic contribution (\ref{nt1}). In contrast, the power-law
contribution (\ref{res2}) dominates the behavior of the imaginary part of
the susceptibility, so that the logarithmic contribution (\ref{nt2}) is
irrelevant.

Let us finally discuss in brief the nonlinear response associated with the
DW motion. We define the nonlinear susceptibilities as $\chi _{m}=\chi
_{m}^{\prime }-i\chi _{m}^{\prime \prime }$ with
\begin{eqnarray}
&&\chi _{m}^{\prime }=\frac{1}{\pi h_{0}}\int\limits_{0}^{2\pi }P(t)\cos
(m\omega t)d(\omega t),  \label{chi1} \\
&&\chi _{m}^{\prime \prime }=\frac{1}{\pi h_{0}}\int\limits_{0}^{2\pi
}P(t)\sin (m\omega t)d(\omega t).  \label{chi2}
\end{eqnarray}%
The linear susceptibility corresponds to $\chi _{1}$. To calculate the
nonlinear susceptibility in the sliding regime, i.e. at very low frequencies
$\omega <\omega ^{\ast }$, we integrate Eq.~(\ref{P1}) with $\langle
v(t)\rangle $ given by Eq.~(\ref{creep}) and $h(t)=h_{0}\cos (\omega t)$,
and then substitute $P(t)$ in Eqs.~(\ref{chi1}) and (\ref{chi2}). \cite%
{brazovskii03} As a result we obtain $\chi _{m}^{\prime }=0$ and $\chi
_{m}^{\prime \prime }\propto 1/\omega $ for odd $m$. Note that in the linear
case ($m=1$) this gives a conduction-type susceptibility $\chi (\omega
)=1/i\omega \tau _{0}$. The computation of the nonlinear susceptibility in
the stochastic regime is a more difficult task which is left for further
investigations.

In conclusion, we have considered the dielectric response of ferroelectric
DWs below the paraelectric-ferroelectric phase transition temperature. As a
function of the frequency of the external electric AC-field there are three
regimes in the behaviour of the complex susceptibility: the sliding, the
stochastic, and the pinned regimes. The response in the sliding regime,
which corresponds to very low frequencies, is due to the creep-like motion
of DWs. The response in the stochastic regime is related to jumps of the DW
as a whole, and can be described using the concept of waiting time
distributions. The response in the pinned regime at high-frequencies is
expected to be due to superposition of center of mass motion and
relaxational motion of internal modes. The existence of all three regimes
for the dielectric response was recently observed in $\mathrm{SrTi}^{18}%
\mathrm{O}_{3}$. \cite{Dec04}

The support from the Deutsche Forschungsgemeinschaft (SFB 418) is gratefully
acknowledged.

\end{document}